\documentclass{article}
\usepackage{amsmath,amssymb}
\usepackage{authblk}
\usepackage{cite}

\makeatletter
\@addtoreset{equation}{section}

\makeatother

\newcommand{\R}{\mathcal R}
\newcommand{\D}{\mathcal D}
\newcommand{\hc}{\mathrm{h.c.}}
\newcommand{\del}{\partial}
\newcommand{\eps}{\varepsilon}
\newcommand{\pphi}{\varphi}
\newcommand{\Tr}{\mathrm{Tr}}
\newcommand{\tr}{\mathrm{tr}}
\newcommand{\Slash}[1]{{\ooalign{\hfil/\hfil\crcr$#1$}}}

\newcommand{\fd}{\tilde{\D}}
\newcommand{\fdsla}{\Slash{\fd}}
\newcommand{\E}{\mathcal E}

\title{\textbf{Quantum Correction from Super-Weyl Transformation in Supergravity}}
\author{Akihisa D.-E. Tateishi}
\affil{\it Department of Physics, The University of Tokyo, Tokyo 113-0033, Japan}
\date{}

\begin{document}
\maketitle

\begin{abstract}
Einstein-frame supergravity is accompanied by quantum correction terms because of super-Weyl transformation. The correction term consists of the field strength terms that respectively originate from gauge, gravitational, and K\"ahler anomalies. In this paper, it is shown how the full-kind one-loop Jacobian correction term is calculated and represented in $\mathcal N=1, D=4$ superspace formalism.
\end{abstract}

\newpage

\section{Introduction}
Various supersymmetric theories have been subject to researches and successful as quantum field theories till today. For example, path integral calculation of supersymmetric Yang-Mills theory has already been deeply researched \cite{sw0, sw00, sw1, sw2, sw3}. However, supergravity theory has not been understood so clearly as super-Yang-Mills, due to the complexity of its Lagrangian and the difficulty in its quantisation. In this paper, as a foothold for a better understanding of quantum supergravity, we aim to calculate the qunatum correction term from super-Weyl transformation in $\mathcal N=1, D=4$ supergravity.
\par
Matter-coupled supergravity Lagrangian is usually constructed via the Brans-Dicke form, where the scalar curvature (i.e. the kinetic term of the graviton) is unconventionally coupled with the matter scalar:
\begin{align}
\mathcal L_{\mathrm{BD}}/e=-\frac12e^{-K/3}\R+\cdots\label{eq:bransdickelag}
\end{align}
and thereafter transformed into the Einstein form by super-Weyl transformation:
\begin{align}
\mathcal L_\mathrm{E}/e=-\frac12\R+\cdots.\label{eq:einsteinlag}
\end{align}
Here $e$ is the vielbein determinant, $\mathcal R$ the (gravitational) scalar curvature, and $K=K(A,A^\dagger)$, an Hermitian function, the K\"ahler potential for the target manifold of the nonlinear sigma model. Since the transformation from (\ref{eq:bransdickelag}) to (\ref{eq:einsteinlag}) includes Weyl rescaling and chiral rotation of the fermions, the full Einstein-frame Lagrangian in quantum theory requires the anomaly correction term:
\begin{align}
\mathcal L_\mathrm{E}^{\mathrm{full}}=\mathcal L_\mathrm{E}+\Delta \mathcal L.
\end{align}
In this paper we calculate $\Delta\mathcal L$ at one loop. A previous research \cite{4} has already given a similar consideration and calculation, but it is incomplete in that
\begin{itemize}
\item they calculate only the gauge anomaly and ignore the gravitational one and the K\"ahler one
\item they ignore the K\"ahler-gauge coupling in the covariant derivative of the chiral fermion in the case 	where the target manifold has a nontrivial K\"ahler metric.
\end{itemize}
Hence this paper gives the full-kind, one-loop $\Delta\mathcal L$ in the correct form, i.e. with all kinds of (gauge, gravitational, and K\"ahler) anomalies and with the correct gauge anomaly term where the K\"alher metric is nontrivial. In order to do this, we also construct the anomaly-calculation method (Fujikawa method) for the gauge bosons and the Rarita-Schwinger fermion, not only for spin-1/2 fermions, which is already well known. In a previous research \cite{ibehari} they treat the path-integral measure in a prudent but complicated way in order to investigate how the anomaly mediated gaugino mass is determined and expressed in terms of superfield, while we simply calculate the anomaly terms caused by super-Weyl transformation.\par
This paper is organised as follows. In Section 2 we review the general super-Weyl transformation and the specific parameter required when the Brans-Dicke form is transformed into the Einstein form. In Section 3 we review the Fujikawa method for spin-1/2 fields and apply it to construct the similar method of spin-1 and spin-3/2 versions. In Section 4, combining them, we calculate the anomaly correction term from super-Weyl transformation and describe it in the terms of the superfield strengths. Section 5 is the summary.

\section{Super-Weyl Transformation}
There are roughly two ways to construct the supergravity Lagrangian. The one is superspace formalism \cite{spsp1, spsp2, 1}, and the other is superconformal tensor calculus\footnote{Actually there is a construction method integrating both methods, which is called superconformal superspace formalism. Refer to \cite{3, 3-1, 3-2, 3-3, 3-4}, for example.} \cite{2-1, 2-2, 2-3, 2-4, 2-5, conf1, conf2, 2}. This paper is mainly described in superspace formalism, and the notation follows \cite{1} unless otherwise specified.\par
Matter-coupled supergravity Lagrangian by superspace formalism is constructed inevitably via the kinetic terms of the form \cite{1}
\begin{align}
\mathcal L_{\mathrm{BD}}=&\int d^2\Theta\,2\E\left[\frac38\left(\bar\D^2-8R\right)e^{-(K+\Gamma)/3}\right.\nonumber\\
&\quad\left.+\frac1{16}H_{(ab)}(\Phi)W^{(a)\alpha}W^{(b)}_\alpha+P(\Phi)\right]+\hc\\
=&e\left[-\frac12e^{-K/3}\R-\frac14e^{-K/3}\eps^{klmn}\left(\bar\psi_k\bar\sigma_l\psi_{mn}-\psi_k\sigma_l\bar\psi_{mn}\right)\right.\nonumber\\
&\quad\left.-\frac i2\left(\chi^i\sigma^m\D_m\bar\chi_i+\bar\chi^i\bar\sigma^m\D_m\chi_i\right)-\del^mA^i\del_mA^\ast_i+\cdots\right],
\end{align}
which has unconventionally matter-coupled kinetic terms (Brans-Dicke form) of the gravitational multiplet and non-K\"ahler kinetic terms of the chiral matter multiplet. Here $H_{(ab)}$ is the gauge kinetic function, $W^{(a)\alpha}$ the gauge field strength\footnote{$W^{(a)\alpha}$ should not be confused with $W^{\alpha\beta\gamma}$, one of the gravitational field strengths.}, where $(a)$ is the index for the adjoint representation, $P$ the superpotential, $K$ the K\"ahler potential\footnote{Throughout this paper, $K$ denotes $K(\Phi,\Phi^\dagger)$ in the context of superfields, while it denotes $K(A,A^\ast)$ in the context of component fields.}, and $\Gamma=\Gamma(\Phi,\Phi^\dagger,V)$ the gauge counterterm, which renders the Lagrangian gauge invariant. In order to transform Lagrangian into the canonical Einstein (and K\"ahler) form, Weyl transformations and some other transformations are required and they are united as super-Weyl transformation. Thus in this section we resummarise the general super-Weyl transformation in the superfield level and its special parameter for the transformation from Brans-Dicke form to Einstein form. The results in this section are quoted from \cite{1} and \cite{4} as a whole.

\subsection{General Super-Weyl Transformation}
Since supergravity in superspace has a large number of redundant degrees of freedom, torsion constraints are required to reduce them. Super-Weyl transformation is the transformation that includes supervielbein rescaling and preserves the torsion constraints. A general infinitesimal super-Weyl transformation is parametrised by a chiral superfield $\Sigma$ as\footnote{These transformation rules should be understood in $\Theta$ expansion: for example, in general $\chi$ in $\Phi$ can be transformed even though $\Phi$ is of Weyl weight zero, since the supervielbein $E_M^A$ is transformed and $\Theta$ expansion is dependent on the supervielbein via the covariant derivative.}
\begin{align}
\delta_{\mathrm{SW}}\E&=6\Sigma\E+\frac\del{\del\Theta^\alpha}(S^\alpha\E)\nonumber\\
\delta_{\mathrm{SW}}\Phi&=-S^\alpha\frac\del{\del\Theta^\alpha}\Phi\nonumber\\
\delta_{\mathrm{SW}}\left[\left(\bar\D^2-8R\right)U\right]&=-(\bar\D^2-8R)\left[(4\Sigma-2\Sigma^\dagger)U\right]\nonumber\\
&\qquad-S^\alpha\frac\del{\del\Theta^\alpha}\left[\left(\bar\D^2-8R\right)U\right]\nonumber\\
\delta_{\mathrm{SW}}W_\alpha&=-3\Sigma W_\alpha-S^\beta\frac\del{\del\Theta^\beta}W_\alpha,\label{eq:infinitesimalsw}
\end{align}
where
\begin{align}
\left.S^\alpha=\Theta^\alpha\left(2\Sigma^\dagger-\Sigma\right)\right|+\Theta^2\left.\left(\D^\alpha\Sigma\right)\right|
\end{align}
and $U$ is an arbitrary real superfield of super-Weyl weight zero\footnote{Here we require only the case of weight zero, but in more general the field $U$ may have nonzero super-Weyl weight. See \cite{1} for that case.}. These transformations induce an infinitesimal variation of the Lagrangian:
\begin{align}
\delta\mathcal L=\int d^2\Theta\,2\E\left\{\frac38\left(\bar\D^2-8R\right)\left[2\left(\Sigma+\Sigma^\dagger\right)e^{-(K+\Gamma)/3}\right]+6\Sigma  P\right\}+\hc
\end{align}
Thus a general finite transformation is
\begin{align}
\mathcal L\mapsto\mathcal L'=\int d^2\Theta\,2\E\left\{\frac38\left(\bar\D^2-8R\right)\exp\left[-\frac13\left(K+\Gamma-6\Sigma-6\Sigma^\dagger\right)\right]\right.\nonumber\\
\left.+\frac14H_{(ab)}W^{(a)}W^{(b)}+e^{6\Sigma}P\right\}+\hc\label{eq:superweyl}
\end{align}

\subsection{Parameters from Brans-Dicke to Einstein}
By transformation (\ref{eq:superweyl}), the Lagrangian can be transformed into the form without undesirable couplings: without Brans-Dicke couplings of the gravitational multiplet or non-K\"ahler kinetic terms of the matter multiplet. In order to operate this transformation, it is necessary that $\Sigma$ should satisfy
\begin{align}
K|=6(\Sigma+\Sigma^\dagger)|,\quad(\D_\alpha K)|=6(\D_\alpha\Sigma)|,\quad(\D^2K)|=6(\D^2\Sigma)|,\label{eq:swparameter}
\end{align}
or $K-6\Sigma-6\Sigma^\dagger$ has no lowest, $\Theta$, and $\Theta^2$ components \cite{4}. The first condition corresponds to the canonical kinetic term of the graviton, the second to that of the gravitino, and the third to those of the matter multiplets\footnote{Refer to Appendix \ref{chap:parameter} for a detailed explanation.}. Note that $\Gamma$ has no contribution to (\ref{eq:swparameter}), since $\Gamma$ can be assumed to be in the Wess-Zumino gauge, where $\Gamma$ has no lowest, $\Theta$, $\Theta^2$ coefficients \cite{1}. These conditions determine the chiral parameter $\Sigma$ required for the transition to Einstein frame:
\begin{align}
\Sigma=B+\sqrt2\Theta\pphi+\Theta^2C,
\end{align}
with
\begin{align}
B=\frac1{12}K+i\phi,\quad\pphi=\frac16K_i\chi^i,\quad C=\frac16K_iF^i-\frac1{12}K_{ij}\chi^i\chi^j,\label{eq:parameterBDtoE}
\end{align}
where
\begin{align}
K_i\equiv\frac{\del K}{\del A^i},\quad K_{ij}\equiv\frac{\del^2K}{\del A^i\del A^j},
\end{align}
and $F^i$ is the $\Theta^2$ component of $\Phi^i$. Note that $\phi$, the imaginary part of the lowest component, is left arbitrary. From the point of view that the Lagrangian should be of the Einstein and K\"ahler form, $\phi$ can be set to 0 by the condition (\ref{eq:swparameter}). However, the degree of freedom of $\phi$ plays a significant role in the anomaly calculation, as shown in Chapter \ref{chap:jacobian}.

\section{Fujikawa Method}
The transformation (\ref{eq:infinitesimalsw}) includes Weyl rescaling of each field,  and furthermore chiral rotations of the fermions. Since these transformations are anomalous, a correction term that originates in the variation of the path integral measure should be added to the Lagrangian in the quantum theory\footnote{The anomaly correction term cannot be avoided even if one adopt the Einstein-frame Lagrangian as the starting point of the quantum theory, since the Einstein-frame SUSY transformation contains Weyl rescaling.}. In order to calculate this correction term, an anomaly calculation technique, which is called Fujikawa method, can be utilised. Originally Fujikawa method was devised for the calculation of symmetry breaking (i.e. how much different the divergence of the conserved current is from zero), but the functional measure variation from the field redifinition can also be calculated in almost the same way. Although Fujikawa method for spin-1/2 fields is already well known \cite{5}, the similar methods for spin-1 and spin-3/2 fields are yet to be clearly constructed. Thus in this section we review Fujikawa method and construct the similar anomaly calculation methods for the gauge vector bosons and the Rarita-Schwinger field.

\subsection{Generalities}
In general, the functional measure $\D\phi$, of a field $\phi$ (not necessarily a scalar), for the (Wick-rotated and Euclideanised) path integral
\begin{align}
\int\D\phi\,e^{-S[\phi]}=\int\D\phi\,\exp\left(-\int d^4x\,\mathcal L[\phi]\right)
\end{align}
is defined as
\begin{align}
\D\phi\equiv\prod_nda_n,
\end{align}
where $a_n$ is the expansion coefficients by the orthonormal basis:
\begin{align}
\phi(x)=\sum_na_n\phi_n(x)\\
\int d^4x\,\phi^\dagger_m(x)\phi_n(x)=\delta_{mn},
\end{align}
and the basis functions $\phi_n(x)$ are the eigenfunctions of the ``Hermite, gauge-covariant differential operator" $\mathcal O$:
\begin{align}
\mathcal O\phi_n=\lambda_n\phi_n.
\end{align}
Note that $\phi_n$ and $a_n$ satisfy
\begin{align}
a_n=\int d^4x\,\phi^\dagger_n(x)\phi(x)\\
\sum_n\phi_n(x)\phi^\dagger_n(y)=\delta(x-y).\label{eq:delta}
\end{align}
Then for an infinitesimal transformation $\phi\mapsto\phi'=\phi+\alpha\phi$ with a parameter $\alpha$,\footnote{Note that in general the parameter $\alpha$ is not necessarily a scalar, but may have spinor or Lorentz indices if the field $\phi$ is a spinor or a vector.} which satisfies $|\alpha|\ll1$,
\begin{align}
a'_n&=\int d^4x\,\phi^\dagger_n(x)\phi'(x)\nonumber\\
&=\sum_ma_m\left(\delta_{mn}+\int d^4x\,\phi^\dagger_n(x)\alpha(x)\phi_m(x)\right)\nonumber\\
&\equiv\sum_ma_m\left(\delta_{mn}+M_{mn}\right),
\end{align}
and the functional measure varies as
\begin{align}
\D\phi'&=\D\phi\,\det\left(\frac{\del a'}{\del a}\right)\\
\det\left(\frac{\del a'}{\del a}\right)&=\exp\tr\log\left(\delta_{mn}+M_{mn}\right)\nonumber\\
&\sim\exp\tr M\nonumber\\
&=\exp\left[\sum_n\int d^4x\,\phi^\dagger_n(x)\alpha(x)\phi_n(x)\right].
\end{align}
However, the sum $\sum_n\phi_n(x)\phi_n(x)$ is ill-defined because of (\ref{eq:delta}). Therefore this sum should be regularised in a gauge-invariant way:
\begin{align}
&\sum_n\int d^4x\,\phi^\dagger_n(x)\alpha(x)\phi_n(x)\nonumber\\
\to&\left.\int d^4x\,\Tr\left[\alpha(x)\sum_ne^{t^2\lambda_n^2}\phi_n(x)\phi^\dagger_n(x)\right]\right|_{t=0},
\end{align}
where Tr denotes the full trace, i.e. the trace with respect to all indices such as spinor or coordinate ones\footnote{Remember that in general $\phi$ and $\alpha$ can have spinor or coordinate indices.}. Furthermore, the internal sum is 
\begin{align}
\sum_ne^{t^2\lambda_n^2}\phi_n(x)\phi^\dagger_n(x)&=\left.\sum_ne^{t^2\lambda_n^2}\phi_n(y)\phi^\dagger_n(x)\right|_{y=x}\nonumber\\
&=\left.e^{t^2\mathcal O_x^2}\sum_n\phi_n(y)\phi_n^\dagger(x)\right|_{y=x}\nonumber\\
&=\left.e^{t^2\mathcal O_x^2}\delta(x-y)\right|_{y=x}\nonumber\\
&=\int\frac{d^4k}{(2\pi)^4}e^{-ikx}e^{t^2\mathcal O^2}e^{ikx}.\label{eq:planewave}
\end{align}
Thus it suffices to define appropriate $\mathcal O$'s and calculate the integral (\ref{eq:planewave}) for spinors, vectors, and Rarita-Schwinger fields.\par
The generalities above can be applied also to theories that contains gravity, but note that the transformation parameter should be for the weight-corrected field $\tilde\phi\equiv\sqrt e\phi$ with no diffeomorphism indices\footnote{In the argument below we use fields with diffeomorphism indices for simplicity in the calculation of the spacetime trace, but the results are not affected. Only the transformation parameters are relevant.}. For example, the parameter for the gravitino should be understood as the one for $\tilde\psi_a\equiv\sqrt e\psi_a$, not for $\psi_m$ or $\psi_a$. Here shown are the coefficients for the infinitesimal transformation (\ref{eq:infinitesimalsw}) with $\Sigma|=\zeta+i\xi$:
\begin{align}
\delta e^a_m&=2\zeta e^a_m\nonumber\\
\delta\tilde\psi_a&=3(\zeta-i\xi)\tilde\psi_a\nonumber\\
\delta\tilde A&=0\nonumber\\
\delta\tilde\chi&=3(\zeta+i\xi)\tilde\chi\nonumber\\
\delta\tilde F&=2(\zeta+3i\xi)\tilde F\nonumber\\
\delta\tilde\lambda&=(\zeta-3i\xi)\tilde\lambda\nonumber\\
\delta\tilde v_a&=-2\zeta\tilde v_a\nonumber\\
\delta\tilde D&=(\zeta-3i\xi)\tilde D.\label{eq:trweight}
\end{align}

\subsection{Spinor Field: Review}
The path integral approach to the quantum anomaly of a spinor has long been established \cite{5}. The operator $\mathcal O$ should be taken to be $\fdsla\equiv\gamma^m\fd_m$, where $\fd_m$ is the full covariant derivative, including the gravitational, gauge, and K\"ahler connections. Then the integral (\ref{eq:planewave}) is estimated as
\begin{align}
&\int\frac{d^4k}{(2\pi)^4}e^{-ikx}\exp\left[t^2\fdsla^2\right]e^{ikx}\nonumber\\
=&\int\frac{d^4k}{(2\pi)^4}\exp\left[t^2\left(\fdsla+i\Slash k\right)^2\right]\nonumber\\
=&\frac1{t^4}\int\frac{d^4k}{(2\pi)^4}\exp\left(-k^2+2it\Slash k\fdsla+t^2\fdsla^2\right)\qquad(kt\mapsto k)\label{eq:heatkernel}\\
\xrightarrow{t\to0}&\frac1{384\pi^2}\left[2\Omega^{mn}\Omega_{mn}+2\left(\Box\gamma^{mn}\Omega_{mn}\right)+3\left(\gamma^{mn}\Omega_{mn}\right)^2\right],\label{eq:spinorgeneral}
\end{align}
where $\Omega_{mn}=[\fd_m,\,\fd_n]$, the field strength of the covariant derivative\footnote{Actually (\ref{eq:heatkernel}) contains terms that diverge as $1/t^4$ and $1/t^2$, but they cancel out those from the functional measure variation of the bosonic field. Refer to Appendix \ref{chap:hke} for a detailed calculation.}.

\subsection{Vector Field}
The path integral measure for a vector field can be defined basically in a similar way to that for a spinor, but the gauge degree of freedom should be treated carefully. A candidate for the differential operator $\mathcal O$ is $\mathcal O^2A_k=\nabla^l\left(\nabla_lA_k-\nabla_kA_l\right)$. However, this $\mathcal O^2$ is not elliptic, and therefore the integral (\ref{eq:planewave}) does not converge. This is because $A_k$ has the gauge freedom, and therefore the integrand of (\ref{eq:planewave}) is constant in the direction of this freedom. Thus it is necessary to fix the gauge and suppress the divergence. In order to do that, the diffeomorphism-invariant Lorentz gauge condition $\nabla_kA^k=0$ can be adopted. Then the second-order differential operator above
\begin{align}
\nabla^l\left(\nabla_lA_k-\nabla_kA_l\right)&=\Box A_k+\R_k^{\phantom kl}A_l\nonumber\\
&\sim\left(\delta_k^{\phantom kl}\Box+\R_k^{\phantom kl}\right)A_l\nonumber\\
&\equiv(\Box+\hat \R)A_k
\end{align}
is now elliptic. The plane-wave trace (\ref{eq:planewave}) is evaluated as
\begin{align}
&\int\frac{d^4k}{(2\pi)^4}e^{-ikx}e^{t^2(\Box+\hat \R)}e^{ikx}\nonumber\\
\xrightarrow{t\to0}&\frac1{192\pi^2}\left[2\left(\Box\R\right)+6\R^{mn}\R_{mn}-\R^{klmn}\R_{klmn}\right]\label{eq:vectortracemeasure}
\end{align}
where $\R_{klmn}$ and $\R_{mn}$ are the (gravitational) Riemann curvature tensor and Ricci tensor respectively. Note that in supergravity theory $A_m$ has only the gravitational connection in its covariant derivative, and therefore the result is written in only terms of the curvature tensor from the metric. Since $A_m$ is transformed only by Weyl rescaling (i.e. scalar-like transformation), we have already taken the coordinate trace in (\ref{eq:vectortracemeasure}).

\subsection{Rarita-Schwinger Field}
A candidate for the gauge-covariant operator is $\mathcal O\psi^k=(1/\sqrt2)\gamma^{klm}\nabla_l\psi_m$, but the gauge condition $\nabla^m\psi_m=0$ needs to be imposed because of the same reason as the case for vector fields. Then $\mathcal O^2$ is modified into an elliptic operator:
\begin{align}
\mathcal O^2\psi^k&=\gamma^{klm}\nabla_l\gamma_{mnp}\nabla^n\psi^p\nonumber\\
&\sim\left[\delta^k_l\Box+X^k_{\phantom kl}\right]\psi^l\nonumber\\
&\equiv(\Box+\hat X)\psi^k
\end{align}
where
\begin{align}
X^k_{\phantom kl}=\frac14\delta^k_l\gamma^{mn}\Omega_{mn}+\frac12\left(\gamma^{km}\Omega_{ml}+\gamma_{lm}\Omega^{mk}\right).
\end{align}
The spacetime trace (\ref{eq:planewave}) is calculated as
\begin{align}
&\int\frac{d^4k}{(2\pi)^4}e^{-ikx}e^{t^2(\Box+\hat X)}e^{ikx}\nonumber\\
\xrightarrow{t\to0}&\frac1{384\pi^2}\left[-3\left(\gamma^{mn}\Omega_{mn}\right)^2+6\gamma^{kl}\Omega_{lm}\gamma^{mn}\Omega_{nk}-16\Omega^{mn}\Omega_{mn}\right]\nonumber\\
=&\frac1{3072\pi^2}\left(6\R^2-12\R^{mn}\R_{mn}+22\R^{klmn}\R_{klmn}-11\gamma^{abcd}\R^{mn}_{\phantom{mn}ab}\R_{mncd}\right).\label{eq:rsmeasure}
\end{align}
Similarly to (\ref{eq:vectortracemeasure}), we have already taken the coordinate trace. The spinor-index trace is yet to be summed.

\section{Jacobian Correction Term\label{chap:jacobian}}
In this section we finally calculate the Jacobian correction term $\Delta\mathcal L$ in one-loop. First, as a preparation, we in advance write down the square terms of the Yang-Mills and gravitational field strengths in superfield notation. Then we compute only the terms in $\Delta\mathcal L$ that can be written as the squares of the bosonic field strengths and describe them in terms of superfield by coefficient comparison.\par
Yang-Mills field strength is
\begin{align}
\int d^2\Theta\,2\E\Sigma W^{(a)\alpha}W^{(a)}_\alpha+\hc=-e\xi V^{(a)mn}\tilde V^{(a)}_{mn}+\cdots,
\end{align}
where $\Sigma$ is a chiral superfield and $\Sigma|=\zeta+i\xi$ and the ellipses denote the terms including fermions.\par
In Poincar\'e supergravity there are three gravitational field strengths \cite{4thsg}\footnote{Note that the reference \cite{4thsg} is incomplete in that they have not calculated the parity-odd term $\epsilon\R\R$ of (\ref{eq:sfstrg1}), which derives from the imaginary part of $W^{\alpha\beta\gamma}W_{\alpha\beta\gamma}$.}:
\begin{align}
&\int d^2\Theta\,2\E\Sigma W^{\alpha\beta\gamma}W_{\alpha\beta\gamma}+\hc\nonumber\\
=&e\left[-\frac1{24}\zeta\left(\R^2-2\R^{mn}\R_{mn}+\R^{klmn}\R_{klmn}\right)+\frac1{48}\xi\epsilon^{abcd}\R^{mn}_{\phantom{mn}ab}\R_{mncd}+\cdots\right]\label{eq:sfstrg1}
\end{align}
\begin{align}
\int d^2\Theta\,2\E\Sigma\left(\bar\D^2-8R\right)\left(G^{\alpha\dot\alpha}G_{\alpha\dot\alpha}\right)+\hc&=-e\zeta\left(2\R^{mn}\R_{mn}-\frac49\R^2+\cdots\right)\\
\int d^2\Theta\,2\E\Sigma\left(\bar\D^2-8R\right)\left(R^\dagger R\right)+\hc&=-e\zeta\cdot\frac1{18}\R^2+\cdots.
\end{align}\par
Furthermore, there is a K\"ahler-curvature field strength of chiral superfield:
\begin{align}
&&\int d^2\Theta\,2\E\Sigma\left(\bar\D^2-8R\right)&\Big[R^{ij^\ast}_{\phantom{ij^\ast}kl^\ast}(\Phi)R_{ij^\ast mn^\ast}(\Phi)\nonumber\\
&&&\quad\times(\D^\alpha\Phi^k)(\bar\D_{\dot\alpha}\Phi^{\dagger l})(\D_\alpha\Phi^m)(\bar\D^{\dot\alpha}\Phi^{\dagger n})\Big]+\hc\nonumber\\
&=&e\zeta\cdot 8R^{ij^\ast}_{\phantom{ij^\ast}kl^\ast}(A)R_{ij^\ast mn^\ast}&(A)(\D^pA^k)(\D^qA^{\dagger l})(\D_pA^m)(\D_qA^{\dagger n})+\cdots,
\end{align}
where $R_{ij^\ast kl^\ast}(\Phi)=R_{ij^\ast kl^\ast}(\Phi,\Phi^\dagger)$ is the curvature tensor of the target K\"ahler manifold in terms of superfield, and $R_{ij^\ast kl^\ast}(A)=R_{ij^\ast kl^\ast}(A,A^\dagger)$ similarly. From here, we operate the calculation of the Jacobians and describe them in terms of the superfield strengths above, for each of chiral, vector, and gravitational multiplets.

\subsection{Chiral Multiplet}
The curvature of covariant derivative (i.e. the commutator of covariant derivative) for the chiral spinor is
\begin{align}
[\D_m,\D_n]\chi^i&=M^{\phantom{mn}i}_{mn\phantom ij}\chi^j+\frac14\R_{mnab}\gamma^{ab}\chi^i,
\end{align}
where
\begin{align}
M^{\phantom{mn}i}_{mn\phantom ij}&=V_{mn}^{(a)}\nabla_jX^{(a)i}+R^i_{\phantom ijkl^\ast}(A)(\D_mA^k)(\D_nA^{\dagger l}),
\end{align}
and $X^{(a)i}$ is the Killing vectors of the target K\"ahler manifold. Thus the plane-wave trace (\ref{eq:spinorgeneral}) is estimated as
\begin{align}
&2\Omega^{mn}\Omega_{mn}+2\left(\Box\gamma^{mn}\Omega_{mn}\right)+3\left(\gamma^{mn}\Omega_{mn}\right)^2\nonumber\\
=&N\left(-\Box\R+\frac34\R^2-\frac14\R^{klmn}\R_{klmn}+\frac18\gamma^{abcd}\R^{mn}_{\phantom{mn}ab}\R_{mncd}\right)\nonumber\\
&\quad+3\gamma^{klmn}V^{(a)}_{kl}V^{(b)}_{mn}(\nabla_iX^{(a)j})(\nabla_jX^{(b)i})\nonumber\\
&\quad+4R^{ij^\ast}_{\phantom{ij^\ast}kl^\ast}(A)R_{ij^\ast mn^\ast}(A)(\D^pA^k)(\D^qA^{\dagger l})(\D_pA^m)(\D_qA^{\dagger n})+\cdots,\label{eq:chiraljcbsum}
\end{align}
where $N$ is the number of chiral multiplets. Finally, using (\ref{eq:trweight}), the one-loop correction term from the chiral multiplets is\footnote{Note that (\ref{eq:chiraljcbsum}) is yet to be summed in spinor index.}
\begin{align}
\Delta\mathcal L^1_{\mathrm{chiral}}=\Delta\mathcal L^1_{\mathrm{chiral;K}}+\Delta\mathcal L^1_{\mathrm{chiral;V}}+\Delta\mathcal L^1_{\mathrm{chiral;G}},
\end{align}
where
\begin{align}
&&\Delta\mathcal L^1_{\mathrm{chiral;K}}\qquad\qquad\qquad\qquad\nonumber\\
=&&\frac1{64\pi^2}\int d^2\Theta\,2\E\Sigma\left(\bar\D^2-8R\right)&\Big[R^{ij^\ast}_{\phantom{ij^\ast}kl^\ast}(\Phi)R_{ij^\ast mn^\ast}(\Phi)\nonumber\\
&&&\quad\times(\D^\alpha\Phi^k)(\bar\D_{\dot\alpha}\Phi^{\dagger l})(\D_\alpha\Phi^m)(\bar\D^{\dot\alpha}\Phi^{\dagger n})\Big]+\hc\label{eq:jchik}
\end{align}
\begin{align}
\Delta\mathcal L^1_{\mathrm{chiral;V}}=-\frac3{16\pi^2}\int d^2\Theta\,2\E\Sigma W^{(a)\alpha}W^{(b)}_\alpha\nabla_iX^{(a)j}\nabla_jX^{(b)i}+\hc\label{eq:jchiv}
\end{align}
\begin{align}
&&\Delta\mathcal L^1_{\mathrm{chiral;G}}\qquad\quad&\nonumber\\
=&&\frac N{128\pi^2}\int d^2\Theta\,2\E\Sigma&\Big[24W^{\alpha\beta\gamma}W_{\alpha\beta\gamma}\nonumber\\
&&&\quad+(\bar\D^2-8R)(G^{\alpha\dot\alpha}G_{\alpha\dot\alpha}-64R^\dagger R+6\D^2R)\Big]+\hc\label{eq:jchig}
\end{align}

\subsection{Vector Multiplet}
The curvature of covariant derivative for the spinor in the vector multiplet is
\begin{align}
[\D_m,\D_n]\lambda^{(a)}=-f^{(abc)}V_{mn}^{(b)}\lambda^{(c)}+\frac14\R_{mnab}\gamma^{ab}\lambda^{(a)},
\end{align}
where $f^{(abc)}$ is the structure constant of the gauge group. Then, together with (\ref{eq:vectortracemeasure}) and using (\ref{eq:trweight}), the correction term from the vector multiplet is evaluated as
\begin{align}
\Delta\mathcal L^1_{\mathrm{vector}}=\Delta\mathcal L^1_{\mathrm{vector;V}}+\Delta\mathcal L^1_{\mathrm{vector;G}},
\end{align}
where
\begin{align}
\Delta\mathcal L^1_{\mathrm{vector;V}}=\frac3{16\pi^2}N_G\int d^2\Theta\,2\E\Sigma W^{(a)\alpha}W^{(a)}_\alpha+\hc\label{eq:jvecv}
\end{align}
\begin{align}
&&\Delta\mathcal L^1_{\mathrm{vector;G}}\qquad\qquad\quad&\nonumber\\
=&&\frac 3{128\pi^2}N_G\int d^2\Theta\,2\E\Sigma&\Big[-8W^{\alpha\beta\gamma}W_{\alpha\beta\gamma}\nonumber\\
&&&\quad+(\bar\D^2-8R)(G^{\alpha\dot\alpha}G_{\alpha\dot\alpha}+8R^\dagger R+2\D^2R)\Big]+\hc\label{eq:jvecg}
\end{align}
Here $N_G$ is the Dynkin index of gauge multiplet, normalised to $n$ for $SU(n)$.

\subsection{Gravitational Multiplet}
The correction term from the gravitational multiplet is directly calculated from (\ref{eq:rsmeasure}) and (\ref{eq:trweight}):
\begin{align}
\Delta\mathcal L^1_{\mathrm{gravity}}=\frac1{16\pi^2}\int d^2\Theta\,2\E\Sigma\Big[&-33W^{\alpha\beta\gamma}W_{\alpha\beta\gamma}\nonumber\\
&+(\bar\D^2-8R)(-G^{\alpha\dot\alpha}G_{\alpha\dot\alpha}+10R^\dagger R)\Big]+\hc\label{eq:jg}
\end{align}

\section{Summary}
In this paper we have calculated the one-loop quantum correction term to the supergravity Lagrangian from super-Weyl field redefinition, including the transformation from the Brans-Dicke form to the Einstein-frame Lagrangian. This correction term is essentially inevitable since the Einstein-frame supersymmetry transformation includes Weyl rescaling and chiral rotation of the fields. The correction term is in total described as
\begin{align}
\Delta\mathcal L^1=&\Delta\mathcal L^1_{\mathrm{chiral;K}}+\Delta\mathcal L^1_{\mathrm{chiral;V}}+\Delta\mathcal L^1_{\mathrm{chiral;G}}\nonumber\\
&+\Delta\mathcal L^1_{\mathrm{vector;V}}+\Delta\mathcal L^1_{\mathrm{vector;G}}+\Delta\mathcal L^1_{\mathrm{gravity}},
\end{align}
where each term is evaluated as (\ref{eq:jchik}), (\ref{eq:jchiv}), (\ref{eq:jchig}), (\ref{eq:jvecv}), (\ref{eq:jvecg}), and (\ref{eq:jg}) in terms of superfield.\par
In order to calculate the Jacobian correction term, we first expanded the anomaly-calculation method, which is already well known as Fujikawa method for a spinor, to the ones for a vector boson and a Rarita-Schwinger field. Thereafter, by coefficient comparison, we adjusted the numerical factors in front of superfield strengths and finally determined the entire form of the correction term.\par
The entire correction term is in a complicated form with many terms and miscellaneous coefficients. This is because we are considering Poincar\'e supergravity. Poincar\'e supergravity is a theory that is obtained by gauge fixing from conformal supergravity, and therefore Poincar\'e supergravity is a ``mathematically halfway'' theory in a sense. On the other hand one could consider an anomaly calculation similar to this paper in conformal supergravity. Since anomaly calculation is a ``purely mathematical'' operation, it might be expected that the anomaly term could be simpler in conformal supergravity, where all the field strengths, including the K\"ahler curvature, are put in a single supermultiplet. Further researches are expected for this topic.

\section*{Acknowledgements}
The author would like to express gratitude to Takeo Moroi for the idea of this research and meaningful discussions.

\appendix
\section{Determination of Super-Weyl Parameter\label{chap:parameter}}
In this section we derive the conditions (\ref{eq:swparameter}), which finally determine the parameters (\ref{eq:parameterBDtoE}), for the super-Weyl transformation from Brans-Dicke form to Einstein form. For the detail of $\Theta$ expansion of each superfield, refer to \cite{1}.\par
First, out of the full Lagrangian
\begin{align}
\int d^2\Theta\,2\E\left(\D^2-8R\right)\exp\left[-\frac13\left(L+\Gamma\right)\right],
\end{align}
where $L=K-6\Sigma-6\Sigma^\dagger$, the kinetic terms of each fields are included in the term
\begin{align}
\int d^2\Theta\,2\E\left(\D^2-8R\right)L.
\end{align}
Since
\begin{align}
\left(2\E R\right)\Big|_{\Theta=0}=\frac16e\R+\cdots,
\end{align}
it is required that $L|=0$ for the Einstein canonical kinetic term of the graviton.\par
Next, since
\begin{align}
\left.\left(\D_\alpha R\right)\right|=-\frac16\sigma_a\bar\sigma_b\psi^{ab}_\alpha+\cdots,
\end{align}
with $\psi_{mn}=\D_m\psi_n-\D_n\psi_m$, it should be imposed that $\left(\D_\alpha L\right)|=0$ in order to remove differential coupling like $\chi\sigma^{ab}\psi_{ab}$.\par
Finally, since the term
\begin{align}
\bar\D^2K(\Phi,\Phi^\dagger)=-\Theta^2K_{i^\ast j^\ast}\D^mA^{\dagger i}\D_mA^{\dagger j}+\cdots
\end{align}
spoils the K\"ahler form of the matter kinetic term, it is necessary to adjust with $\Sigma$ so that $(\D^2L)|=0$.

\section{Calculation of Heat Kernel Expansion\label{chap:hke}}
In this section we derive a formula in order to calculate the spacetime trace (\ref{eq:spinorgeneral}), (\ref{eq:vectortracemeasure}), and (\ref{eq:rsmeasure}). The methods and results in this section are in reference to \cite{HKex} as a whole.\par
First,
\begin{align}
&\int\frac{d^4k}{(2\pi)^4}e^{-ikx}e^{t^2(\Box+\hat X)}e^{ikx}\nonumber\\
=&\int\frac{d^4k}{(2\pi)^4}\exp\left\{t^2\left[\left(\D+ik\right)^2+\hat X\right]\right\}\\
=&\frac1{t^4}\int\frac{d^4k}{(2\pi)^4}\exp\left[t^2\left(\Box+\hat X\right)+2itk\cdot\D-k^2\right]\\
\sim&\int\frac{d^4k}{(2\pi)^4}\frac1{24}e^{-k^2}\times\nonumber\\
&\quad:\left[16(k\cdot\D)^4+12(-4)\left(\Box+\hat X\right)(k\cdot\D)^2+12\left(\Box+\hat X\right)^2\right]:,\label{eq:HeatKernelGeneral}
\end{align}
where $\sim$ denotes extracting the constant term of the Laurent expansion in $t$, the colon $:X:$ denotes Weyl-ordered product of $X$, i.e. 
\begin{align}
:AB^2:\ =\frac13\left(AB^2+BAB+B^2A\right),
\end{align}
and we have used
\begin{align}
\frac1{t^4}e^{At^2+Bt+C}\sim\frac1{24}:e^C\left(B^4+12AB^2+12A^2\right):.
\end{align}
Continue to calculate (\ref{eq:HeatKernelGeneral}) to obtain
\begin{align}
&\int\frac{d^4k}{(2\pi)^4}e^{-ikx}e^{t^2(\Box+\hat X)}e^{ikx}\nonumber\\
=&\frac1{96\pi^2}\Big\{\Box^2+\D^{\mu}\Box\D_{\mu}+\D^{\mu}\D^{\nu}\D_{\mu}\D_{\nu}\nonumber\\
&\quad\quad\quad-2\left[\left(\Box+\hat X\right)\Box+\D^{\mu}\left(\Box+\hat X\right)\D_{\mu}+\left(\Box+\hat X\right)\Box\right]\nonumber\\
&\quad\quad\quad+3\left(\Box+\hat X\right)^2\Big\}\\
=&\frac1{96\pi^2}\left[\D^{\mu}\D^{\nu}\D_{\mu}\D_{\nu}-\D^{\mu}\Box\D_{\mu}+3\hat X^2+\Box\hat X-2\D^{\mu}\hat X\D_{\mu}+\hat X\Box\right]\label{eq:b12}\\
=&\frac1{96\pi^2}\left[\frac12\Omega^{\mu\nu}\Omega_{\mu\nu}+3\hat X^2+\left(\Box\Hat X\right)\right],\label{eq:b13}
\end{align}
\footnote{The term $\Box\hat X$ in the expression (\ref{eq:b12}) should be understood as a differential operator, i.e. $\Box\hat X(f)\equiv\Box(\hat Xf)$, while $(\Box\hat X)$ in (\ref{eq:b13}) as a mere multiplication operator.}where $\Omega_{\mu\nu}\equiv\left[\D_{\mu},\D_{\nu}\right]$ and we have used
\begin{align}
\int d^4k\,e^{-k^2}&=\pi^2\\
\int d^4k\,e^{-k^2}k_{\mu}k_{\nu}&=\frac{\pi^2}2g_{\mu\nu}\\
\int d^4k\,e^{-k^2}k_{\mu}k_{\nu}k_{\rho}k_{\sigma}&=\frac{\pi^2}4\left(g_{\mu\nu}g_{\rho\sigma}+g_{\mu\rho}g_{\nu\sigma}+g_{\mu\sigma}g_{\nu\rho}\right).
\end{align}
For example, in the case for a spinor,
\begin{align}
\Box+\hat X=\fdsla^2=\gamma^{\mu}\gamma^{\nu}\fd_{\mu}\fd_{\nu}=\Box+\frac12\gamma^{\mu\nu}\Omega_{\mu\nu},
\end{align}
and therefore $\hat X\equiv\frac12\gamma^{\mu\nu}\Omega_{\mu\nu}$. Substitute this into (\ref{eq:b13}) and finally obtain (\ref{eq:spinorgeneral}).

\end{document}